\documentclass{article}
\usepackage{epsf}
\title{Variational Methods in Loop Quantum Cosmology}
\author{Fatimah Shojai \and Ali Shojai
\\                  
  Department of Physics, University of Tehran,
Tehran, Iran\\
  and \\
  Max Planck Institute for Gravitational Physics, Golm, Potsdam, Germany
}
\begin{document}

\maketitle

\begin{abstract}
An action functional for the loop quantum cosmology difference
equation is presented. It is shown that by guessing the
general form of the solution and optimizing the action
functional with respect to the parameters in the guessed
solution one can obtain approximate solutions which are 
reasonably good.
\end{abstract}

\section{Introduction}
One of the main candidates for quantum gravity is loop quantum gravity which is canonical quantization of gravity using connection  variables. It is background independent and non-perturbative. This last property makes it suitable for considering quantum effects on black holes and big bang.
Applying the theory to cosmology leads to loop quantum cosmology (LQC)\cite{martin} whose main equation for the evolution of the scale factor, that is the 
Hamiltonian constraint, is a difference equation.
This difference equation may be solved numerically\cite{green}, or
exactly in some cases\cite{ss}. Also continuous approximation, in which
the difference equation is approximated by a differential
equation, maybe useful at least for some parts of the domain
of the difference equation\cite{martin,martin1}.
The most important result of LQC is the resolution of big bang singularity\cite{martin2}. Also there are phenomenological results from the effective equations that incorporates the modiffications of matter hamiltonian due to quantum gravity( see e.g. \cite{martin3})

Here we shall study the possibility of adopting approximate
methods based on variational methods for solving LQC difference equation. First we
shall briefly review the variational methods for difference
equations\cite{logan}. Then, we shall write an action functional
appropriate for the difference equation of LQC and obtain the first integrals of the equation.
Next we shall find approximations to the solutions, not by solving the
difference equation but by guessing a solution and optimizing
the action, and compare these approximate solutions with the
exact ones.
\section{Review of variational methods for difference equations}
Here we summarize the results of ref. \cite{logan} for a case suitable for our problem. Suppose that we have an action of the form:
\begin{equation}
{\cal A}=\sum_{n=M_1}^{M_2}{\cal L}(n;Q_n,Q_{n-1},Q_{n-2};Q^*_n,Q^*_{n-1},Q^*_{n-2}))
\end{equation}
in which $Q_n$ is the degree of freedom of the system with Lagrangian ${\cal L}$. The equation of motion is derived using the least action principle stating that the action should be stationary with respect to the variations
$Q_n\rightarrow Q_n+\delta Q_n$.
The variation in the Lagrangian is:
\begin{equation}
\delta {\cal L}(\delta Q_n,\delta Q^*_n)={\cal R}^{(1)}\delta Q_n +{\cal R}^{(2)}\delta Q^*_n
\end{equation}
where the operators ${\cal R}^{(1)}$ and ${\cal R}^{(2)}$ are defined as:
\begin{equation}
{\cal R}^{(1)}=\sum_{k=0}^2 \partial_{Q_{n-k}}{\cal L}(n;Q_n,Q_{n-1},Q_{n-2};Q^*_n,Q^*_{n-1},Q^*_{n-2})(L^{(1)})^k
\end{equation}
\begin{equation}
{\cal R}^{(2)}=\sum_{k=0}^2 \partial_{Q^*_{n-k}}{\cal L}(n;Q_n,Q_{n-1},Q_{n-2};Q^*_n,Q^*_{n-1},Q^*_{n-2})(L^{(2)})^k
\end{equation}
and the lowering operators $L^{(1)}$ and $L^{(2)}$ are defined as:
\begin{equation}
L^{(1)}X(Q_n)=X(Q_{n-1});\ \ \ and\ \ \ L^{(2)}X(Q^*_n)=X(Q^*_{n-1})
\end{equation}
Defining the adjoint operators:
\begin{equation}
\overline{\cal R}^{(1)}=\sum_{k=0}^2 \partial_{Q_{n-2+k}}
{\cal L}(n+2-k;Q_{n+2-k},Q_{n+1-k},Q_{n-k};Q^*_{n+2-k},Q^*_{n+1-k},Q^*_{n-k})(L^{(1)})^k
\end{equation}
\begin{equation}
\overline{\cal R}^{(2)}=\sum_{k=0}^2 \partial_{Q^*_{n-2+k}}
{\cal L}(n+2-k;Q_{n+2-k},Q_{n+1-k},Q_{n-k};Q^*_{n+2-k},Q^*_{n+1-k},Q^*_{n-k})(L^{(2)})^k
\end{equation}
and using the Lagrange identity proved in ref. \cite{logan}:
\begin{equation}
{\cal R}^{(1)}\delta Q_n=\delta Q_n \overline{\cal R}^{(1)} 1 + \Delta B^{(1)}(\delta Q_n,Q^*_n);\ \ and \ \ {\cal R}^{(2)}\delta Q^*_n=\delta Q^*_n \overline{\cal R}^{(2)} 1 + \Delta B^{(2)}(\delta Q_n,Q^*_n)
\end{equation}
where $\Delta$ means evaluation of the difference of the quantity at $n-1$ and $n$ and with:
\[ B^{(1)}(\delta Q_n,\delta Q^*_n)=-\sum_{i=1}^2\sum_{j=1}^i\delta Q_{n-j}\partial_{Q_{n-j}} \]
\begin{equation}
{\cal L}(n-j+i;Q_{n-j+i},Q_{n-j+i-1},Q_{n-j+i-2},Q^*_{n-j+i},Q^*_{n-j+i-1},Q^*_{n-j+i-2})
\end{equation}
\[ B^{(2)}(\delta Q_n,\delta Q^*_n)=-\sum_{i=1}^2\sum_{j=1}^i\delta Q^*_{n-j}\partial_{Q^*_{n-j}} \]
\begin{equation}
{\cal L}(n-j+i;Q_{n-j+i},Q_{n-j+i-1},Q_{n-j+i-2},Q^*_{n-j+i},Q^*_{n-j+i-1},Q^*_{n-j+i-2})
\end{equation}
and substituting in the action, one sees that the $\Delta$ terms are zero by not varying the boundary (initial) conditions and the ${\cal R}$ adjoint terms leads to the equation of motion as:
\begin{equation}
\sum_{k=0}^2\partial_{Q_n}{\cal L}(n+2-k;Q_{n+2-k},Q_{n+1-k},Q_{n-k},Q^*_{n+2-k},Q^*_{n+1-k},Q^*_{n-k})=0
\end{equation}
and its complex conjugate.

In order to find first integrals of the above difference equation, one first should define a symmetry of the Lagrangian. A Lagrangian is said to be symmetric under the transformation:
\begin{equation}
Q_n\rightarrow Q_n+\epsilon u^{(1)}(n;Q_n,Q^*_n);\ \ \ \ \ Q^*_n\rightarrow Q^*_n+\epsilon u^{(2)}(n;Q_n,Q^*_n)
\end{equation}
provided one can find some $v(n;Q_n,Q_{n-1},Q_{n-2},Q^*_n,Q^*_{n-1},Q^*_{n-2})$ for which:
\begin{equation}
\delta {\cal L}=\epsilon\Delta v(n;Q_n,Q_{n-1},Q_{n-2},Q^*_n,Q^*_{n-1},Q^*_{n-2})
\end{equation}
If a Lagrangian is symmetric in this way, we have a constant of motion as:
\begin{equation}
J=\epsilon v(n;Q_n,Q_{n-1},Q_{n-2},Q^*_n,Q^*_{n-1},Q^*_{n-2})
-\sum_{i=1}^2 B^{(i)}(u^{(i)}(n;Q_n,Q^*_n))
\end{equation}
\section{Variational principle and first integrals for LQC}
The suitable Lagrangian for LQC difference equation is:
\begin{equation}
{\cal L}=\kappa^{-1}|Q_n-Q_{n-2}|^2+\left ( \kappa^{-1}(\Omega-2)+S_{n-1}\right )|Q_{n-1}|^2
\label{last1}
\end{equation}
leading to the well--known equations of motion:
\begin{equation}
Q_{n-2}-\Omega Q_n+Q_{n+2}=\kappa S_nQ_n
\label{last2}
\end{equation}
where $S_n$ is the matter contribution to the difference equation.\footnote{The quantum dynamics of LQC states is obtained from the Hamiltonian constraint difference equation:
\[
(V_{\ell+5\ell_0}-V_{\ell+3\ell_0})e^{-2i\Gamma\ell_0}\psi(\phi,\ell+4\ell_0)
-\Omega(V_{\ell+\ell_0}-V_{\ell-\ell_0})\psi(\phi,\ell)+
\]
\[
(V_{\ell-3\ell_0}-V_{\ell-5\ell_0})e^{2i\Gamma\ell_0}\psi(\phi,\ell-4\ell_0)=
-\frac{1}{3} \kappa\gamma^3\ell_0^3\ell_p^2\hat{C}_{m}^\mu(\ell,\phi)\psi(\phi , \ell)
\]
in which $\Omega=2-4\ell_0^2\gamma^2\Gamma (\Gamma-1)$, $\hat{C}_{m}$ is the matter ($\phi$) Hamiltonian, and  $\ell_p$  is the Planck length. $\Gamma$ is the spin connection parameter defined as $\Gamma=k/2$, so $\Omega=2$ for a flat model and $\Omega=2+\ell_0^2\gamma^2>2$ for a closed model. $\ell_0$ and $\mu$ are quantum ambiguities, $\gamma$ is Immirizi--Barbero parameter, $\kappa=8\pi G$, and $V_\ell=\left ( \frac{ \gamma |\ell|}{6}\right )^{3/2}\ell_p^3$ are
the eigenvalues of the volume operator. Changing to the variable $Q(\ell)=(V_{\ell+\ell_0}-V_{\ell-\ell_0})e^{-i\ell\Gamma/2}e^{-i\tilde{\omega}\phi}\psi(\phi,\ell)$
and choosing $\ell=2\ell_0 n$ leads to the difference equation mentioned in the text. Although this discussion is about a massless scalar field, it is not hard to see that other matter sources as well as cosmological constant leads to the same form of difference equation.}

Let us now investigate the first integrals. First consider the case $\Omega=2$ and $S_n=0$. Then the Lagrangian has translational symmetry generated by $u^{(i)}=1$.
This leads to $v=0$. The corresponding conserved quantity is:
\begin{equation}
J=-(Q_{n+1}+Q_n-Q_{n-1}-Q_{n-2}+Q^*_{n+1}+Q^*_n-Q^*_{n-1}-Q^*_{n-2})
\end{equation}
which can be simply checked that it is conserved. For this case the equation of motion can be solved as $Q_n=(A+Bn)e^{ip\pi n}=\textit{constant}$ where $p$ is an integer, and $A$ and $B$ are constants\cite{ss}. It can be checked that $J=\textit{constant}$.

A more useful result can be obtained when one considers the complete Lagrangian (with matter and curvature contributions). The Lagrangian has global phase transformation invariance. That is to say, under transformations $Q_n\rightarrow Q_ne^{i\alpha}$
or in its infinitesimal form $Q_n\rightarrow Q_n+i\alpha Q_n$,
the corresponding $v$ is zero, and the conserved quantity can be found as:
\begin{equation}
J=i\alpha(Q^*_{n+1}Q_{n-1}-Q_{n+1}Q^*_{n-1}+Q^*_nQ_{n-2}-Q_nQ^*_{n-2})
\end{equation}
This conserved quantity can give some information about the relation between the real part and imaginary parts of $Q$. Assuming $Q_n=p_n+iq_n$ one shall get $X_n+X_{n-1}=C_1$
where $C_1$ is a constant, and $X_n=p_{n+1}q_{n-1}-p_{n-1}q_{n+1}$.
The solution for $X_n$ is $X_n=-\frac{C_1}{2}+(-1)^{n-1}C_2$
where $C_2$ is another constant. Writing $q_n=p_n Y_n$ we have $Y_{n+2}-Y_n=\frac{C_2(-1)^n-C_1/2}{p_{n+2}p_n}$
which can be solved for $Y_n$ leading to the following relation between the real and imaginary parts of $Q_n$:
\begin{equation}
\frac{q_n}{p_n}=C_3+C_4(-1)^n+\frac{1}{2}\sum_{j=0}^{n-2}(1+(-1)^{n-j})\frac{C_2(-1)^j-C_1/2}{p_jp_{j+2}}
\end{equation}
where $C_3$ and $C_4$ are two other constants.
\section{Solving the difference equation}
The variational principle can also be used for deriving approximate solutions for cases in which one has not an exact solution. This can be done by guessing a solution and choosing the parameters such that the action takes its extremum value. We shall illustrate this method with some examples for which we have the exact solutions.
\section{Case: $\Omega=2$ and $S_n=0$}
The action is:
\begin{equation}
{\cal A}=\kappa^{-1}\sum_{n=-\infty}^{\infty}|Q_n-Q_{n-2}|^2
\end{equation}
One can guess a solution of type $e^{\alpha n}$. Substituting in the action we have:
\begin{equation}
{\cal A}=4\kappa^{-1}\sinh^2\alpha \sum_{n=-\infty}^{\infty}e^{(\alpha+\alpha^*)n}
\end{equation}
Minimizing this leads to either $\alpha=0$ or $\alpha=ip\pi$. Both are exact solutions. A more general guess for $Q$ is $n^\beta e^{\alpha n}$, leading to action:
\begin{equation}
{\cal A}=\kappa^{-1}\sum_{n=-\infty}^{\infty}e^{(\alpha+\alpha^*)n}|e^\alpha (n+1)^\beta e^{-\alpha}(n-1)^\beta|^2
\end{equation}
Minimizing this leads to either $\alpha=ip\pi$, $\beta=0$ for which the solution is $Q_n=\textit{constant}$, or $\alpha=ip\pi$, $\beta=1$ where the solution becomes $Q_n=n$. Both are exact solutions\cite{ss}.
\section{Case: $\Omega\ne 2$ and $S_n=0$}
\begin{figure}[htb]
\begin{center}
\epsffile{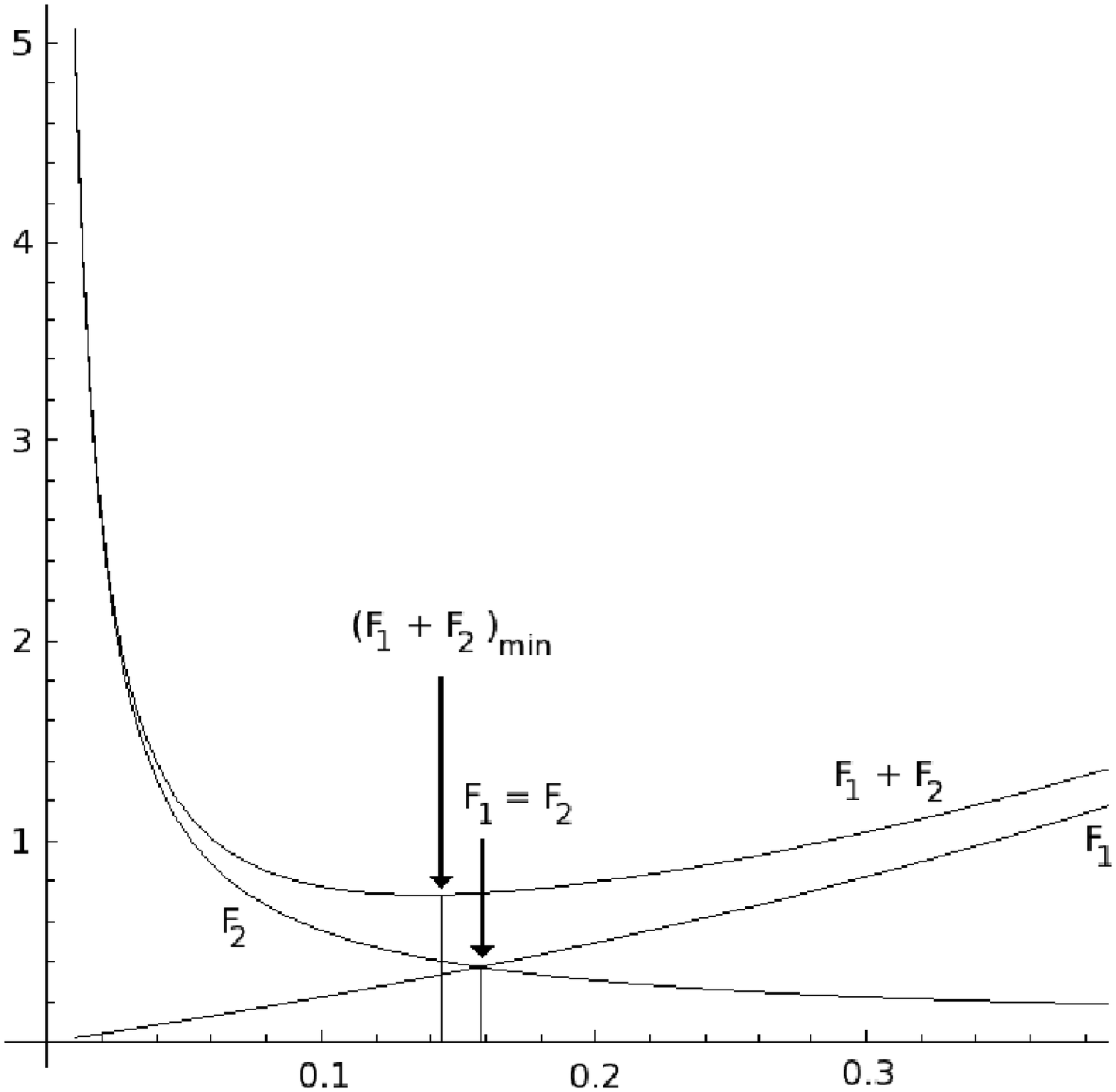}
\end{center}
\caption{Comparison of the minimum of function $F_1+F_2$ and the point $F_1=F_2$.}
\label{fig}
\end{figure}

Now the action would be:
\begin{equation}
{\cal A}=\kappa^{-1}\sum_{n=-\infty}^{\infty}\left ( |Q_n-Q_{n-2}|^2+(\Omega-2)|Q_{n-1}|^2\right )
\end{equation}
Guessing a solution of the form $e^{\beta n}$, we have:
\begin{equation}
\textit{first term}=\kappa^{-1}\sum_{n=-\infty}^{\infty}4\sinh^2 \beta e^{(\beta+\beta^*)n}
\label{x1}
\end{equation}
\begin{equation}
\textit{second term}=\kappa^{-1}\sum_{n=-\infty}^{\infty}(\Omega -2) e^{(\beta+\beta^*)n}
\label{x2}
\end{equation}
The problem is that these summations are divergent, so first we have to make them meaningful by changing the summation limits to $(-N\cdots N)$. Let's assume
$\beta$ is real and positive. Then we have
\begin{equation}
{\cal A}=\kappa^{-1}\left \{ 4\sinh^2\beta +\Omega -2\right \}\frac{e^{-2N\beta}-e^{2\beta(N+1)}}{1-e^{2\beta}}
\end{equation}
Since $N$ should goes to infinity we have ${\cal A}= \kappa^{-1}e^{2\beta N} (F_1(\beta)+F_2(\beta))$
where $F_1(\beta)=e^{2\beta}-1$ and $F_2(\beta)=(\Omega-2)\frac{e^{2\beta}}{e^{2\beta}-1}$.
So the action is (apart from a positive diverging term) sum of two positive terms, one increasing (with respect to $\beta$) and one decreasing, as it is seen in the
figure (\ref{fig}). The minimum value is achieved when one puts this terms nearly equal. 
\epsfxsize=5in
\epsfysize=5in

The minimization of the action, therefore is done by letting $4\sinh^2\beta\sim (\Omega-2)$ leading to $\beta=\pm\beta_0$ with $\beta_0=\frac{1}{2}\cosh^{-1}\frac{\Omega}{2}$.
So the solution is $Q_n=e^{\pm\beta_0 n}$, which is again the exact solution\cite{ss}.

As a result of introducing a cut--off, the minimization depends on it. That is, if we translate the cut--off point from $N$ to $N+c$, the minimum point moves from the point where the derivative of $F_1+F_2$ is zero to the point where  the derivative of $F_1+F_2$ equals to $-2c(F_1+F_2)$.
The solution is to renormalize the regularized action via introducing the physical gravitational constant as: $\kappa_{physical}^{-1}=\kappa^{-1}e^{2\beta N}$.
\section{Case: $\Omega\ne 2$ and $S_n\ne 0$}
Now the action would be:
\begin{equation}
{\cal A}=\sum_{n=-\infty}^{\infty}\left ( \kappa^{-1}|Q_n-Q_{n-2}|^2+(\kappa^{-1}(\Omega-2)+S_{n-1})|Q_{n-1}|^2\right )
\end{equation}
Guessing a solution of type $e^{g(n)}$, we have:
\begin{equation}
\textit{first term}=\kappa^{-1}\sum_{n=-\infty}^{\infty}|e^{g(n)}-e^{g(n-2)}|^2\sim\kappa^{-1}\sum_{n=-\infty}^{\infty}4\sinh^2 g' e^{g+g^*}
\end{equation}
\begin{equation}
\textit{second term}=\sum_{n=-\infty}^{\infty}(\kappa^{-1}(\Omega -2)-S_n) e^{g+g^*}
\end{equation}
Again we make the action extremum by letting these two terms equal.
To see why this is true, we note that the Lagrangian (\ref{last1}) consists of two terms ${\cal L}_1=\kappa^{-1}|Q_{n+1}-Q_{n-1}|^2$ and ${\cal L}_2=\kappa^{-1}(\Omega-2)|Q_n|^2+S_n|Q_n|^2$ (where we have shifted $n$ by one). On the equation of motion (\ref{last2}) the second term would be ${\cal L}_2=\kappa^{-1}Q_n^*(Q_{n+2}+Q_{n-2})-2\kappa^{-1}|Q_n|^2$. Expanding $g(n\pm 1)$ and $g(n\pm 2)$ around $g(n)$ we have ${\cal L}_1\sim \kappa^{-1}e^{g+g^*}|e^{g'}-e^{-g'}|^2$ and ${\cal L}_2\sim \kappa^{-1}e^{g+g^*}(e^{2g'}+e^{-2g'}-2)$. These are equal provided we assume either $g$ is real or purely imaginary. Note that this is an approximate result since we expanded $g$ and choose only the first term of expansions. 

The condition that these two terms be of the same order leads to:
\begin{equation}
g=\pm\int dn \sinh^{-1}\sqrt{\frac{\Omega-2-\kappa S_n}{4}}
\label{last3}
\end{equation}
which is clear that is either real or purely imaginary as we assumed. 
Since we had expanded the first term, we have to have $g''<<g'$ for usability of this solution.
For $S_n=n^\alpha$ with $\alpha$ an arbitrary number, this condition holds. 
\section{cosmological constant}
As an example let us find the solution for the special case in which the matter contribution is cosmological constant. The source term is:
\begin{equation}
S_n=\left \{ \begin{array}{ll}
\Lambda n & |n|>> 1\\
\Lambda |n|^{1/2} & |n|\sim 1
\end{array} \right .
\end{equation}
where $\Lambda$ is proportional to the cosmological constant.
\begin{itemize}
\item $|n|\sim 1$ In this case we have:
\[ g(n)=\pm\frac{-2}{15\Lambda^2\kappa^2}(\Omega-2+\kappa\Lambda|n|^{1/2})^{3/2}(2(\Omega-2)-3\kappa\Lambda|n|^{1/2}) \]
\begin{equation}
\simeq \frac{4(\Omega-2)^{5/2}}{15\Lambda^2\kappa^2}\pm\frac{(\Omega-2)^{1/2}}{2}|n|
\end{equation}
which is very close to exact solution\cite{rej}.
\item $|n|>> 1$ In this case we have:
\[ g(n)=\frac{8}{\kappa\Lambda}\left \{ \frac{1}{2}\left ( \frac{\Omega-2+\kappa\Lambda n}{4}\right )\sinh^{-1}\sqrt{\frac{\Omega-2+\kappa\Lambda n}{4}} \right . \]
\begin{equation}
\left . -\frac{1}{4}\sqrt{\frac{\Omega-2+\kappa\Lambda n}{4}}\sqrt{1+\frac{\Omega-2+\kappa\Lambda n}{4}}+\frac{1}{4}\sinh^{-1}\sqrt{\frac{\Omega-2+\kappa\Lambda n}{4}}\right \}
\end{equation}
Using the asymptotic forms of Bessel functions one can see that this solution is in agreement with the exact ones, which are $J_{n/2+1/\Lambda}(1/\Lambda)$ and $Y_{n/2+1/\Lambda}(1/\Lambda)$\cite{rej}.
\end{itemize}
\section{Other matters}
In general, matter sources in large or small $n$ limit lead to a source term
like $S_n=\Lambda n^\alpha$. For different values of $\alpha$ the integral for
$g(n)$ can be calculated:
\begin{itemize}
\item $\alpha=-2$ (large $n$, massless matter)
\[
g(n)=\frac{n}{\sqrt{\frac{\Omega -1}{\kappa\Lambda }}
   \left(\left(\Omega ^2-3 \Omega +2\right) n^4+\kappa\Lambda 
   (2 \Omega -3) n^2+\kappa\Lambda ^2\right)}
   \times
   \]
\[
\left \{ \sqrt{\frac{\Omega -1}{\kappa\Lambda }}
   \left(\left(\Omega ^2-3 \Omega +2\right) n^4+\kappa\Lambda 
   (2 \Omega -3) n^2+\kappa\Lambda ^2\right) \sinh
   ^{-1}\left(\sqrt{\frac{\kappa\Lambda }{n^2}+\Omega
   -2}\right)\right .
\]
\[
  -i n \kappa\Lambda  \sqrt{\frac{(\Omega -2)
   n^2+\kappa\Lambda }{\kappa\Lambda }} \sqrt{\frac{(\Omega -1)
   n^2+\kappa\Lambda }{\kappa\Lambda }} \sqrt{\frac{\kappa\Lambda
   }{n^2}+\Omega -2} \sqrt{\frac{\kappa\Lambda }{n^2}+\Omega
   -1}\times 
   \]
\begin{equation}
  \left . F\left(i \sinh ^{-1}\left(n \sqrt{\frac{\Omega
   -1}{\kappa\Lambda }}\right);\frac{\Omega -2}{\Omega
   -1}\right)\right \}
\end{equation}
where $F$ is the elliptic function defined as: $F(\phi;m)=\int^\phi_0d\theta (1-m\sin^2\theta)^{-1/2}$.
\item $\alpha=1/2$ (small $n$, massless matter with the quantum ambiguity $\mu=0$) is the same as small $n$, cosmological constant.
\item $\alpha=1$ (small $n$, massless matter with $\mu=1/4$) is the same as large $n$, cosmological constant.
\end{itemize}
\section{Continuous Approximation}
The continuous approximation can be simply achieved by expanding any $Q$ around
$Q_n$ up to second order. The equation of motion is
\begin{equation}
4Q''(n)=(\kappa S(n)+\Omega-2)Q(n)+ {\cal O}(3)
\end{equation}
and the conserved quantity emerging from the phase invariance is 
\begin{equation}
J=\Im (Q(n)Q^{*'}(n))+ {\cal O}(3)
\end{equation}
This is the conserved current expected from the above equation of motion.
\section{Conclusion}
We saw that it is possible to study the LQC difference equation using action functional formulation. We have shown that one can get acceptable solutions by guessing one and making the action optimized. The results are in agreement with numerical and other analytical methods in the literature (see, e.g. \cite{green,rej,a1,a2}). Therefore the method presented here is useful for obtaining analytical solutions in different regimes and for different matter sources.

{\bf Acknowledgments}
The authors like to thank Martin Bojowald, for very fruitful discussions. This work is supported by a grant from University of Tehran.


\begin{thebibliography}{0}

\bibitem{martin}
For a review and complete bibliography see: Bojowald M.,Living Rev. Relativity (Available online at: 
http://www.livingreviews.org/lrr-2005-11), 8, 2005, 11.

\bibitem{green}
Green D. \and Unruh W., Phys. Rev. D, 70, 2004, 103502.

\bibitem{ss}
Shojai A. \and Shojai F., EuroPhys. Lett., 71(6), 2005, 886.

\bibitem{martin1}
Bojowald M., Class. Quant. Grav., 18, 2005, L109.

\bibitem{martin2}
Bojowald M., Phys. Rev. Lett., 86, 2001, 5227.

\bibitem{martin3}
Bojowald M., Phys. Rev. Lett., 89, 2002, 261301.

\bibitem{logan}
Logan J.D., AEQ. Math., 9, 1973, 210.

\bibitem{rej}
Bojowald M. \and Rej A., Class. Quant. Grav., 22, 2005, 3399.

\bibitem{a1}
Ashtekar A., Pawlowski T. \and Singh P., Phys. Rev. D, 73, 2006, 124038.

\bibitem{a2}
Ashtekar A., Pawlowski T. \and Singh P., arXive: gr-qc/0602086, 2006.

\end{thebibliography}
\end{document}